\documentclass[%
 reprint,
 amsmath,amssymb,
 aps,
]{revtex4-2}

\usepackage{graphicx}
\usepackage{here}
\usepackage{dcolumn}
\usepackage{bm}
\usepackage{comment}
\usepackage{parskip}
\usepackage{xcolor}

\begin{document}

\preprint{APS/123-QED}

\title{New type of chaotic solutions found in Gravity model of network transport}

\author{Hajime Koike}
 \email{hajimekoike@outlook.jp}
\author{Hideki Takayasu}
\author{Misako Takayasu}
 \email{corresponding author: takayasu.m.aa@m.titech.ac.jp}
 \affiliation{
  School of Computing, Institute of Science Tokyo, Yokohama 226-8502, Japan
 }
\date{\today}

\begin{abstract}
The gravity model is a mathematical model that applies Newton's universal law of gravitation to socio-economic transport phenomena and has been widely used to describe world trade, intercity traffic flows, and business transactions for more than several decades. However, its strong nonlinearity and diverse network topology make a theoretical analysis difficult, and only a short history of studies on its stability exist. In this study, the stability of gravity models defined on networks with few nodes is analyzed in detail using numerical simulations. It was found that, other than the previously known transition of stationary solutions from a unique diffusion solution to multiple localized solutions, parameter regions exist where periodic solutions with the same repeated motions and chaotic solutions with no periods are realized. The smallest network with chaotic solutions was found to be a ring with seven nodes, which produced a new type of chaotic solution in the form of a mixture of right and left periodic solutions. 
\end{abstract}

\keywords{Chaos, networks, nonlinear transport}
\maketitle


Transport in socio-economic systems is a ubiquitous object, ranging in various spatiotemporal scales, such as movement and exchanges of money and goods, measured at the scale of intra-cities, inter-cities, or countries. Typically, the gravity law is applied to model fluxes in these transport systems such as world trade\cite{tinbergen1963shaping, anderson2011gravity}, population movement\cite{ravenstein1889laws, zipf1946p, lewer2008gravity, park2018generalized, prieto2018gravity}, traffic\cite{erlander1990gravity, jung2008gravity, wang2019gravity, li2021gravity}, human mobility\cite{masucci2013gravity, kwon2023multiple}, and business transaction\cite{tamura2012estimation}. 

Owing to this ubiquity, the gravity law has been a long-standing research topic in social science and economics. A number of works have been conducted on the empirical validation of the gravity law\cite{zipf1946p, lewer2008gravity, park2018generalized, prieto2018gravity, jung2008gravity, wang2019gravity, li2021gravity, masucci2013gravity, kwon2023multiple, tamura2012estimation} or deriving the gravity law theoretically\cite{anderson2011gravity, wang2021free, de2024modelling}, whereas very few studies have been conducted on the stability of transport systems driven by the gravity law. Nevertheless, the stability of the transport systems governed by the gravity law remains a profound subject, because the aforementioned examples of such systems are fundamental components of our society. Studies on the stability of the transport systems are desired both from theoretical understanding of it and practical need to control it.

To address this problem explicitly, one approach involves interpreting the transport by the gravity law as nonlinear transport on networks. A transport system can be regarded as a network, because each site may heterogeneously connect with other sites. The gravity law defined on the edges of the network serves as a nonlinear function to determine the interaction flux, because the flux depends not only on the size of the source site but also on the size of the target site. Along this direction, a model known as ``gravity interaction model"\cite{tamura2018diffusion, koike2022diffusion, koike2024bifurcation} played a role as a working model of open system of gravity law transport, in investigating the role of combination of network topology and nonlinearity on the stability of the state. 

To clarify the role of gravity interaction on stability, a basic question arises: What is the possible nonlinear behavior, such as oscillatory states and chaotic states? The transition from a diffusive state to a localized state is reported in Refs. \cite{tamura2018diffusion, koike2022diffusion}. Recently, our model study revealed hysteresis in a three-node system of gravity law transport\cite{koike2024bifurcation}. These studies focused only on the stability of the fixed points; however, possibilities exist for other types of solutions.

In this letter, we study the gravity interaction model, and show the phase diagram of the model on rings, one of the simplest network structures for investigating behaviors beyond fixed points, to explore diverse solutions. In particular, we found a chaotic region in the localized phase for the first time, even in small networks such as ring networks with only seven nodes. We show strange attractors in such cases, and elucidate the mechanism of the chaotic solution from the perspective of the stability of symmetric localized fixed points. We find that such chaotic behavior occurs for $n \geq 7$ nodes in the ring network, and finally discuss the role of dissipation.


\begin{figure*}[!hbt]
  \centering
  \includegraphics[width=\textwidth, height=0.4\textwidth]{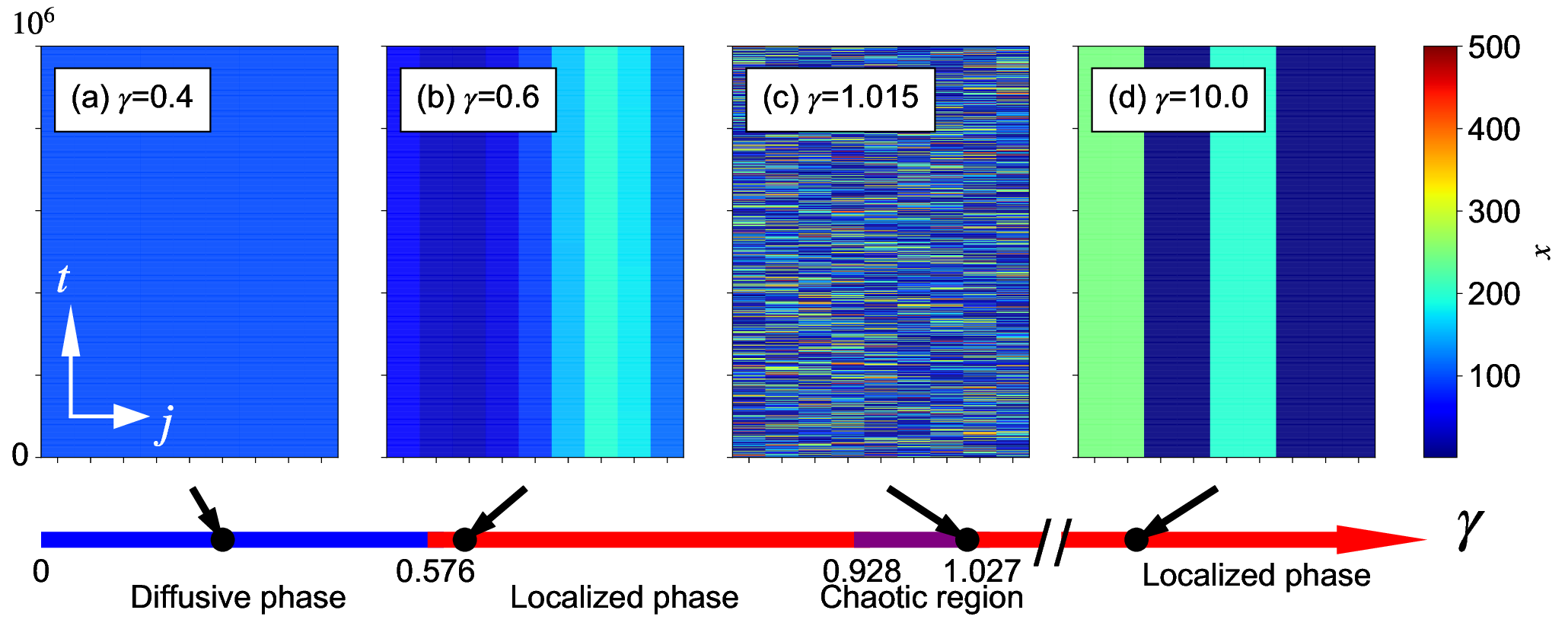}
  \caption{Four phases of our model. Panels (a)-(d) show spatiotemporal patterns of the model's solution of the model, corresponding to four typical states. Horizontal and vertical axes represent space $j$ (node positions aligned in a circular order of ring network) and time $t$, respectively. Color indicates the state variable $x$. (a) Diffusive phase: The fixed point at $\gamma = 0$ is stable. (b) First localized phase: The fixed point at $\gamma = 0$ becomes unstable, and multiple localized fixed points emerge. (c) Chaos region: The trajectory becomes chaotic. (d) Second localized phase: Multiple localized fixed points coexist. }
  \label{Fig:1}
\end{figure*}

The gravity law states that the flux $f_{ij}$ from two adjacent nodes $i$ to $j$ is proportional to $S_i^{\alpha} S_j^{\beta} / D_{ij}^{\delta}$, for some fixed parameters $\alpha, \beta, \delta \geq 0$, where $S_i, S_j$ are the sizes of $i$ and $j$, respectively, and $D_{ij}$ is the distances between $i$ and $j$. The typical value of $\gamma = \beta / \alpha$ is one, as in the case of population movement and world trade. Of course, there are more complicated ways to model fluxes (see e.g., Ref. \cite{kwon2023multiple}), we use this type of gravity law as a simple and typical case of nonlinearity. For simplicity, we consider a ring network, where each node follows a simple identical external interaction coupled with gravity interaction. Because the typical value of distance exponent $\delta$ is as large as either one or two, it is relevant to consider short-range interaction systems as an approximation, not only as a theoretical playground. It is also natural to assume bidirectional interactions bacause interactions do not only comprise only unidirectional links in reality. Then, it is possible to formulate the time evolution of the size variables $x_j = S_j^{\alpha}$ for each node $j=1, 2, \cdots, n$ as follows:
\begin{widetext}
  \centering
  \begin{equation}
    \dot{x}_j = \frac{x_j^\gamma}{x_{j-2}^\gamma + x_j^\gamma} x_{j-1} + \frac{x_j^\gamma}{x_j^\gamma + x_{j+2}^\gamma} x_{j+1} -  x_j - \nu x_j + 1, \; j=1, 2, \cdots, n, 
  \label{Eq:1}
  \end{equation}
\end{widetext}
where $\gamma$ denotes the parameter of nonlinearity of the flux and $\nu$ denotes the dissipation coefficient controlling the external interaction with the outside of the network. The first three terms are couplings, and the last two terms are external intearctions, that is, dissipation and injection. It is possible to recover the typical diffusion at $\gamma=0$. For a more general formulation of the model, see Refs. \cite{tamura2018diffusion, koike2022diffusion, koike2024bifurcation}. Because all elements are identical, $\langle x(t) \rangle \equiv \frac{1}{n} \sum_j x_j(t)$ asymptotically converges to $\nu^{-1}$, implying that the dynamics are confined to a plane in the phase space. In particular, the model \eqref{Eq:1} admits the uniform state $x_j = \nu^{-1} \; \forall j$ for any $\gamma$. Because we consider a ring network, our model has rotational and reflection symmetries.  
For the numerical integration, we used the fourth-order Runge-Kutta method with a fixed time step size of 0.01. The initial state is uniformly sampled from $[\nu^{-1}-1, \nu^{-1}+1]$ in the case $\nu < 1$, and $[\nu^{-1} - 0.1, \nu^{-1} + 0.1]$ in the case $\nu \geq 1$, for each node of the ring network. 

The numerical result reveals that four phases exist in our model: (a) diffusive phase: the fixed point at $\gamma=0$(the uniform state $x_j = \nu^{-1}\; j=1, 2, \cdots, n$) is stable, (b) first localized phase: the fixed point at $\gamma=0$ becomes unstable and multiple localized fixed points become stable, (c) chaotic region: the localized fixed points change their stability and the periodic or chaotic trajectory becomes the attractor, and (d) second localized phase: the localized fixed points become stable again. These four phases in Eq. \eqref{Eq:1} are shown in Fig. \ref{Fig:1}. The four panels shows $10^{6}$ time-unit trajectory. The bifurcation point $\gamma_c$ of the uniform state has been studied as diffusion-localization transition point\cite{tamura2018diffusion, koike2022diffusion}, and is approximately 0.598 for $\nu = 0.01$. The wave-like localized fixed points become stable at this bifurcation. Multiple localized fixed points exist owing to symmetry, and the state eventually converges to one of them depending on the initial state. In the localized phase, the wavelength of the localized fixed point is reduced by increasing $\gamma$. At certain values of $\gamma$, the fixed points become saddles, and periodic trajectory traveling around these saddles become the attractor. This periodic trajectory become chaos at the left onset of the chaotic region. At the right onset of the chaos, the chaotic state become transient chaos, and eventually, the localized fixed points become stable again. In this second localized phase, the localized state consists of two waves of wavelength four, and in each wave there are two peak nodes (see Fig. \ref{Fig:1}(d)). In this state, a main-stream structure\cite{tamura2015extraction} of flow exists; the two nodes at the peak of the wave are sending each other, and the other nodes are sending to the only one of adjacent nodes which is nearer to the peak nodes. No further wavelength shortening occurs by increasing $\gamma$, because this state is stable by strong nonlinearity. The stability of the uniform state presents the characteristic wavelength as a function of $\gamma$, which confirms the above picture\footnote{The eigenvalues of the Jacobian of the model \eqref{Eq:1} at uniform state are exactly calculated in Ref. \cite{koike2022diffusion}, and by taking continuum limit $n \rightarrow \infty$ allows us to discuss the dispersion relation. See Section 1 of the Supplementary Material for details.}.

\begin{figure}[!htb]
  \centering
  \includegraphics[width=\linewidth, height=0.8\linewidth]{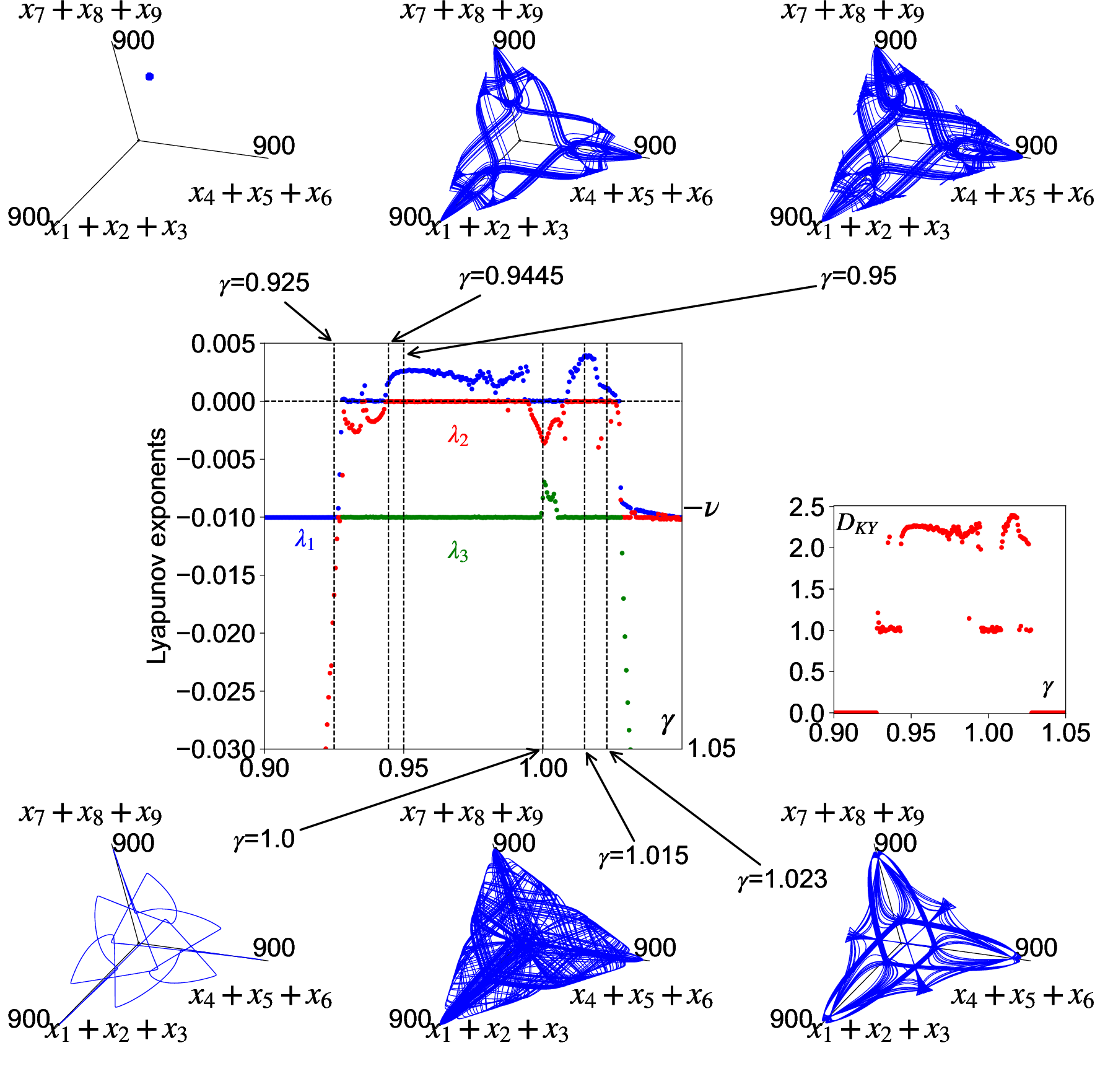}
  \caption{Enlargement of the chaotic region. The center left panel shows top-three Lyapunov exponents $\lambda_1$(blue), $\lambda_2$(red), $\lambda_3$(green) versus $\gamma$, whereas the center right panel shows the Kaplan-Yorke dimension $D_{KY}$ calculated from the Lyapunov spectrum. The top and bottom panels show attractors for different values of $\gamma$ across the chaotic region. Dissipation $\nu$ is fixed to 0.01. Attractors are represented in three-dimensional space, with coordinate axes being $x_1 + x_2 + x_3$, $x_4 + x_5 + x_6$, $x_7 + x_8 + x_9$. The top three attractors are obtained for $\gamma = 0.925$(fixed point attractor), $0.9445$(strange attractor), $0.95$(strange attractor), and the bottom three attractors are obtained for $\gamma = 1.0$(torus attractor), $\gamma = 1.015$(strange attractor), $1.023$(strange attractor). The value of $D_{KY}$ is up to 2.5 throughout chaotic region, justifying the representation of attractors in three-dimensional space.}
  \label{Fig:2}
\end{figure}

\begin{figure}[!htb]
  \centering
  \includegraphics[width=\linewidth, height=\linewidth]{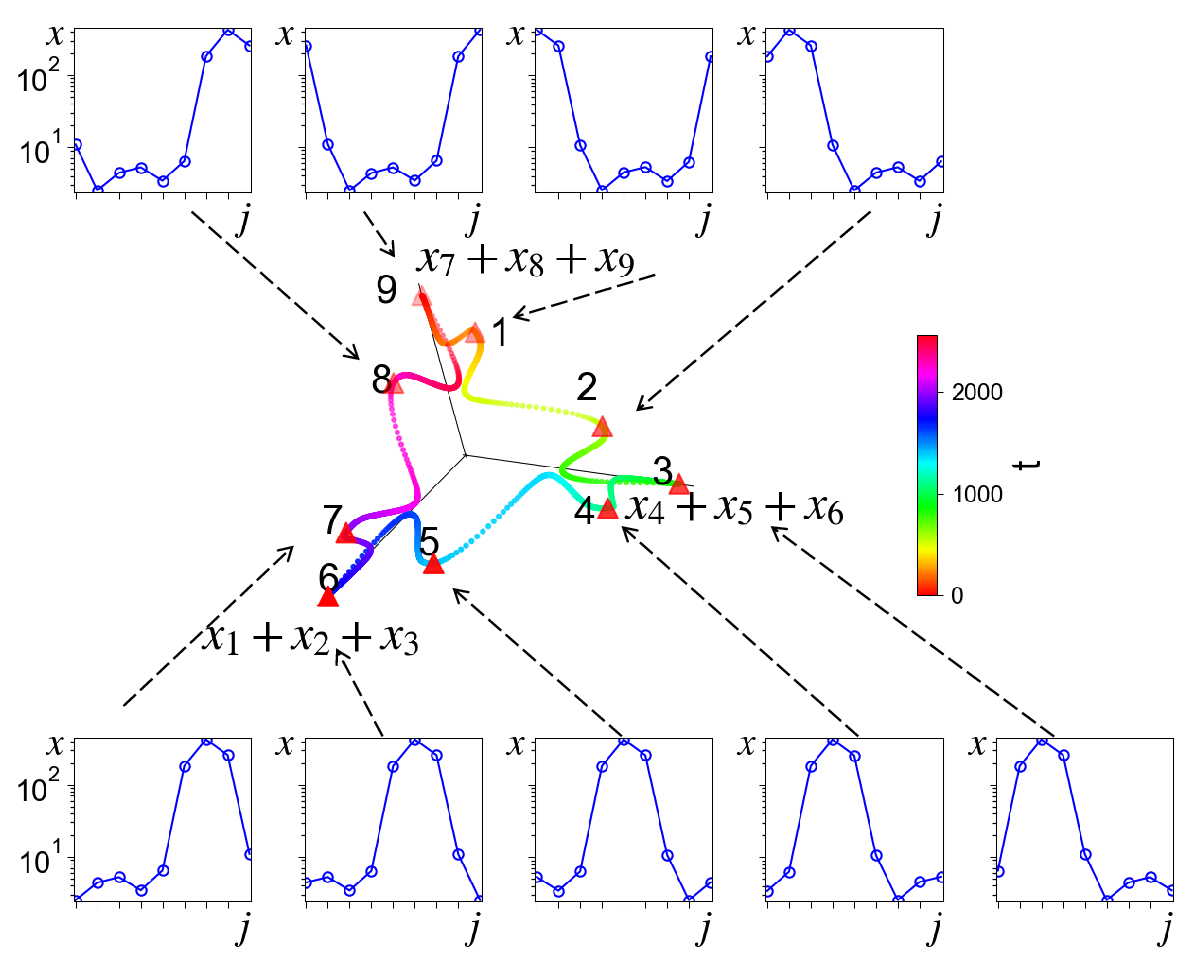}
  \caption{Example of the attractor structure, obtained for $\gamma = 0.943$, in the vicinity of the left onset of chaos. Nine-dimensional trajectory is represented in three-dimensional space in the same mannar as attractors in Fig. \ref{Fig:2}. The trajectory is traveling around nine saddle points (shown in the top and bottom panels, with horizontal and vertical axes being space $j$ and value $x$ respectively) from 1, 2, $\cdots$, 9 in the center panel.}
  \label{Fig:3}
\end{figure}

\begin{figure}[!htb]
  \centering
  \includegraphics[width=\linewidth]{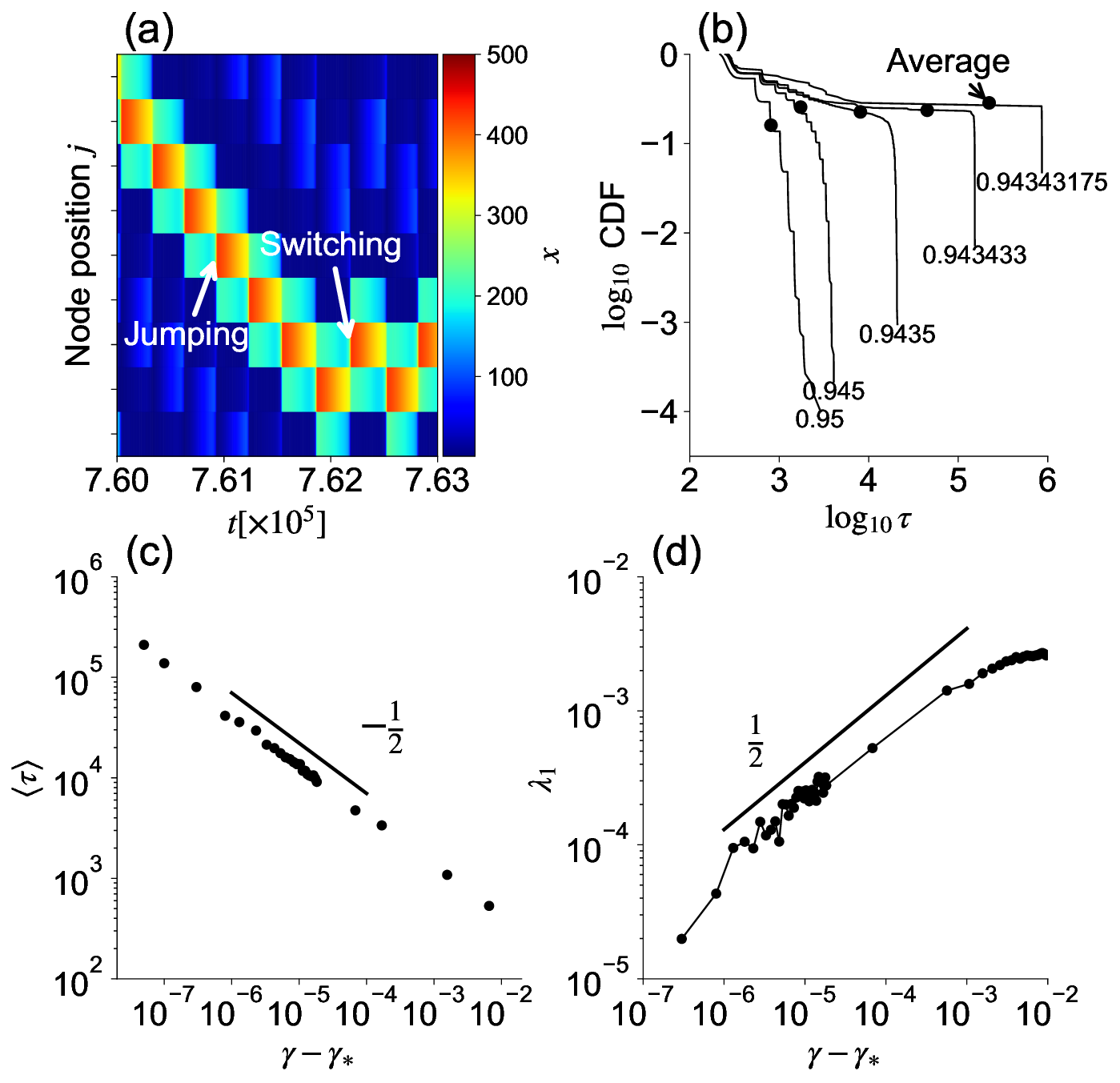}
  \caption{Summary of intermittent feature at the onset of chaos. For $\gamma < \gamma_{\ast} = 0.9434317$, two periodic attractors traveling in opposite directions exists, as shown in Fig. \ref{Fig:3}. At the onset of chaos $\gamma \approx \gamma_{\ast}$, we found intermittent switching direction of traveling. (a) Trajectory at $\gamma = 0.94343175$, slightly above $\gamma_{\ast}$, represented as spatiotemporal plot, as in Fig. \ref{Fig:1}. The terms ``jumping" of peak position, and ``switching" of jumping direction are annotated. (b) Distribution of switching duration $\tau$, measured from single $10^{7}$ time unit trajectory. (c) Average of switching duration. (d) Largest lyapunov exponents $\lambda_1$. Here, Lyapunov exponent is calculated over $10^{6}$ time unit trajectory after $10^{6}$ time unit transient. The scalings of $\langle \tau$ and $\lambda_1$ found here are frequently reported in intermittency route to chaos. }
  \label{Fig:4}
\end{figure}

For the chaotic region, we calculated the Lyapunov exponents numerically. The calculation is based on Shimada-Nagashima algorithm\cite{shimada1979numerical}, with 1 time-unit reorthogonalization, for $10^{5}$ time-unit trajectory after $10^{3}$ time-unit transient. The results are shown in the middle left panel of Fig. \ref{Fig:2}. The results did not change qualitatively for different transient and trajectory time lengths. This revealed the existence of chaotic region for parameters ranging approximately from 0.928 to 1.027. The top and bottom panels show the attractors represented in three-dimensional space, with the axes being the sum of three adjacent nodes, namely, $x_1 + x_2 + x_3$, $x_4 + x_5 + x_6$, and $x_7 + x_8 + x_9$. Kaplan-Yorke dimension $D_{KY}$ is calculated from Lyapunov spectrum, and the result is shown in the middle right panel of Fig. \ref{Fig:2}. $D_{KY}$ is 2.5 at most throughout the chaotic region, which justifies the three-dimensional representation of the attractors. The fixed point attractor($\gamma = 0.925$)\footnote{Only one fixed point is shown in Fig. \ref{Fig:2}, but there are other fixed points owing to symmetry.} becomes saddles, and the attractor becomes a torus($\gamma = 0.943$) connecting these saddles. To illustrate the structure of the strange attractor in the chaotic region, we first show the torus attractor at the left onset of chaos in Fig. \ref{Fig:3}. Three-dimensional representation of attractor is by the same as that in top and bottom panels of Fig. \ref{Fig:2}. The periodic trajectory travels around nine saddles labeled 1, 2, $\cdots$, 9, in this order. This behavior is similar to that of the replicator dynamics, in which invariant subspaces of the model induce structually stable heteroclinic cycles\cite{hofbauer1998evolutionary, chawanya1995new}. One difference is the time spent near the saddles, which is linearly increasing for heteroclinic cycles\cite{may1975nonlinear}, whereas it is constant in our case. Note that the two one-dimensional torus attractors traveling in two directions coexist by reflection symmetry. For a slightly larger value of $\gamma$, we obtained a strange attractor by merging these two tori ($\gamma=0.9445$). At the onset of chaos, switching in traveling direction occurs intermittently. The statistics are summarized in Fig. \ref{Fig:4}. The average duration of switching traveling directions [Fig. \ref{Fig:4}(c)] and the largest Lyapunov exponent [Fig. \ref{Fig:4}(d)] support the common scaling law reported in standard literature of intermittency\cite{pomeau1980intermittent, elaskar2023review}. This strongly suggests intermittency route to chaos in this system.

Once such strange attractor has been formed, it become thick ($\gamma = 0.95$). A region exists near $\gamma=1$ where the attractor becomes a torus again, connecting nine saddle points of different orders from $\gamma=0.943$. For a larger $\gamma$, a region of chaos exists, again, where the attractor is fully developed ($\gamma=1.015$; see Fig. \ref{Fig:1}(c) for spatiotemporal plot). As $\gamma$ reached the right onset of chaos, the attractor changed form ($\gamma=1.023$). In this region, the saddle points differ from those observed in $\gamma=0.943$ and $\gamma=1.0$ cases. At the right onset of chaos, the saddle points found in $\gamma=1.023$ stabilized, and the state was in the second localized phase.

\begin{figure}
  \centering
  \includegraphics[width=\linewidth]{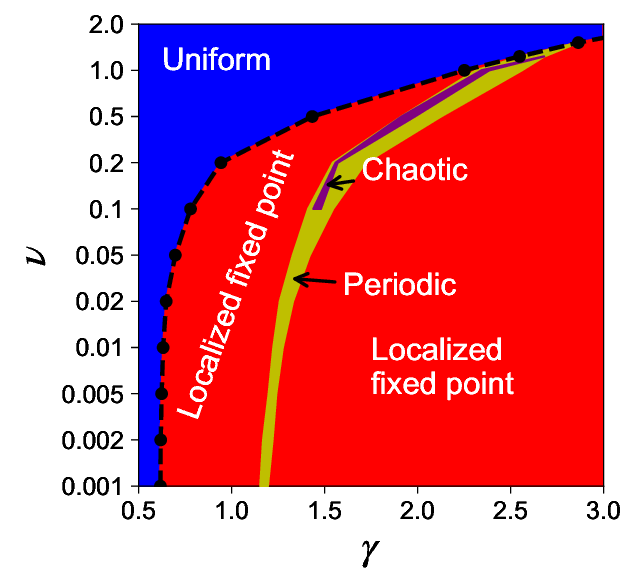}
  \caption{Phase diagram of the model on ring with $n=7$ nodes. Note that Figs. \ref{Fig:1}-\ref{Fig:4} are the results for $n=9$ ring with $\nu=0.01$. The black dashed lines indicate the bifurcation point of the uniform state, which is analytically derived in Ref. \cite{koike2022diffusion}. For small and large $\nu$ values, such as 0.001 and 2.0, the chaotic region disappears. Chaotic region does not exist even outside of the range of $\gamma$ in this figure.}
  \label{Fig:5}
\end{figure}

We have discussed chaos in $n=9$ ring with fixed dissipation $\nu = 0.01$. A similar analysis was performed as shown in Fig. \ref{Fig:2} for rings of different sizes and different values of $\nu$, and found that the minimum number of nodes that exihibit chaos is 7, in the case of ring. The rings with nodes $n \leq 6$ did not exhibit chaotic or periodic solutions. The phase diagram of the model \eqref{Eq:1} for $n=7$ ring is shown in Fig. \ref{Fig:5}. Four regions exist; the uniform state[blue region in Fig. \ref{Fig:5}], which is identical to ``diffusive phase" in large networks\cite{tamura2018diffusion}, two regions of localized fixed points[red region in Fig. \ref{Fig:5}], where the localized pattern stablilizes itself, periodic[yellow region in Fig. \ref{Fig:5}] and chaotic regions[purple region in Fig. \ref{Fig:5}], where the localized patterns are moving regularly or irregularly. One finding is that chaos exists only at intermediate values of dissipation $\nu$, approximately between 0.1 and 1.0. One possible mechanism is as follows: For a sufficiently large dissipation, the dynamics is dominated by external interactions; thus, the uniform state can be stable for high $\gamma$. The localization wavelength immediately after the bifurcation of the uniform state is sufficiently short, such that the localized pattern can be stable. For a suficiently small dissipation, the peak of the localized pattern has almost all the total sizes. The remaining nodes are so small such that no driving forces destabilizing the pattern exist.  

Our analysis shows that the gravity law transport model exhibits chaos in the ring topology; without any heterogeneity of the parameters or complex network topologies. In particular, the chaotic region found in the localized phase, and the system exhibits low-dimensional chaos after the traveling wave-like periodic solution connecting the saddles, owing to the intermittency of the changing traveling direction. These results may be relevant to chaos in social sciences and economics\cite{puu1991nonlinear, puu2013attractors, barnett2015nonlinear, san2020introduction, gordon1992chaos}. Our results could be a bridge between separately discussed subjects in socio-economic systems, namely, gravity law and chaos. Our result highlight several open problems. First, a natural question is the minimal degrees of freedom that exhibit chaos for any network topology, as mentioned previously. We confirmed that none of the strongly connected graphs with four and five nodes exhibit chaotic behaviors. Another open problem is the behavior of the large systems, as in reality. It would also be interesting to study the behavior of the heterogeneous version of our model, for instance, by introducing link weights, distributed dissipation, and a complex functional form of nonlinearity. For example, link weights can be useful to capture the role of distances. The role of such heterogeneity in the stability of a system is a key to understanding the stability of real systems. Last but not least, gravity interaction model successfully reproduced the distribution of company sales\cite{tamura2018diffusion} using the real business network and the parameters estimated from real money flow data. More data collection and analysis of real systems to explore the potential applications of nonlinear transport, along with model studies, are strongly desired for a better understanding of this ubiquitous phenomenon.

This work is supported by a Grant-in-Aid for Scientific Research No. 23KJ0921. HK was financially supported by Japan Society for the Promotion of Science, Research Fellowship for Young Scientists.  

\bibliography{koike_chaos}

\providecommand{\noopsort}[1]{}\providecommand{\singleletter}[1]{#1}%
\begin{thebibliography}{33}%
\makeatletter
\providecommand \@ifxundefined [1]{%
 \@ifx{#1\undefined}
}%
\providecommand \@ifnum [1]{%
 \ifnum #1\expandafter \@firstoftwo
 \else \expandafter \@secondoftwo
 \fi
}%
\providecommand \@ifx [1]{%
 \ifx #1\expandafter \@firstoftwo
 \else \expandafter \@secondoftwo
 \fi
}%
\providecommand \natexlab [1]{#1}%
\providecommand \enquote  [1]{``#1''}%
\providecommand \bibnamefont  [1]{#1}%
\providecommand \bibfnamefont [1]{#1}%
\providecommand \citenamefont [1]{#1}%
\providecommand \href@noop [0]{\@secondoftwo}%
\providecommand \href [0]{\begingroup \@sanitize@url \@href}%
\providecommand \@href[1]{\@@startlink{#1}\@@href}%
\providecommand \@@href[1]{\endgroup#1\@@endlink}%
\providecommand \@sanitize@url [0]{\catcode `\\12\catcode `\$12\catcode
  `\&12\catcode `\#12\catcode `\^12\catcode `\_12\catcode `\%12\relax}%
\providecommand \@@startlink[1]{}%
\providecommand \@@endlink[0]{}%
\providecommand \url  [0]{\begingroup\@sanitize@url \@url }%
\providecommand \@url [1]{\endgroup\@href {#1}{\urlprefix }}%
\providecommand \urlprefix  [0]{URL }%
\providecommand \Eprint [0]{\href }%
\providecommand \doibase [0]{https://doi.org/}%
\providecommand \selectlanguage [0]{\@gobble}%
\providecommand \bibinfo  [0]{\@secondoftwo}%
\providecommand \bibfield  [0]{\@secondoftwo}%
\providecommand \translation [1]{[#1]}%
\providecommand \BibitemOpen [0]{}%
\providecommand \bibitemStop [0]{}%
\providecommand \bibitemNoStop [0]{.\EOS\space}%
\providecommand \EOS [0]{\spacefactor3000\relax}%
\providecommand \BibitemShut  [1]{\csname bibitem#1\endcsname}%
\let\auto@bib@innerbib\@empty
\bibitem [{\citenamefont {Tinbergen}(1963)}]{tinbergen1963shaping}%
  \BibitemOpen
  \bibfield  {author} {\bibinfo {author} {\bibfnamefont {J.}~\bibnamefont
  {Tinbergen}},\ }\bibfield  {title} {\bibinfo {title} {Shaping the world
  economy},\ }\href@noop {} {\bibfield  {journal} {\bibinfo  {journal} {The
  International Executive}\ }\textbf {\bibinfo {volume} {5}},\ \bibinfo {pages}
  {27} (\bibinfo {year} {1963})}\BibitemShut {NoStop}%
\bibitem [{\citenamefont {Anderson}(2011)}]{anderson2011gravity}%
  \BibitemOpen
  \bibfield  {author} {\bibinfo {author} {\bibfnamefont {J.~E.}\ \bibnamefont
  {Anderson}},\ }\bibfield  {title} {\bibinfo {title} {The gravity model},\
  }\href@noop {} {\bibfield  {journal} {\bibinfo  {journal} {Annu. Rev. Econ.}\
  }\textbf {\bibinfo {volume} {3}},\ \bibinfo {pages} {133} (\bibinfo {year}
  {2011})}\BibitemShut {NoStop}%
\bibitem [{\citenamefont {Ravenstein}(1889)}]{ravenstein1889laws}%
  \BibitemOpen
  \bibfield  {author} {\bibinfo {author} {\bibfnamefont {E.~G.}\ \bibnamefont
  {Ravenstein}},\ }\bibfield  {title} {\bibinfo {title} {The laws of
  migration},\ }\href@noop {} {\bibfield  {journal} {\bibinfo  {journal}
  {Journal of the royal statistical society}\ }\textbf {\bibinfo {volume}
  {52}},\ \bibinfo {pages} {241} (\bibinfo {year} {1889})}\BibitemShut
  {NoStop}%
\bibitem [{\citenamefont {Zipf}(1946)}]{zipf1946p}%
  \BibitemOpen
  \bibfield  {author} {\bibinfo {author} {\bibfnamefont {G.~K.}\ \bibnamefont
  {Zipf}},\ }\bibfield  {title} {\bibinfo {title} {The p 1 p 2/d hypothesis: on
  the intercity movement of persons},\ }\href@noop {} {\bibfield  {journal}
  {\bibinfo  {journal} {American sociological review}\ }\textbf {\bibinfo
  {volume} {11}},\ \bibinfo {pages} {677} (\bibinfo {year} {1946})}\BibitemShut
  {NoStop}%
\bibitem [{\citenamefont {Lewer}\ and\ \citenamefont {Van~den
  Berg}(2008)}]{lewer2008gravity}%
  \BibitemOpen
  \bibfield  {author} {\bibinfo {author} {\bibfnamefont {J.~J.}\ \bibnamefont
  {Lewer}}\ and\ \bibinfo {author} {\bibfnamefont {H.}~\bibnamefont {Van~den
  Berg}},\ }\bibfield  {title} {\bibinfo {title} {A gravity model of
  immigration},\ }\href@noop {} {\bibfield  {journal} {\bibinfo  {journal}
  {Economics letters}\ }\textbf {\bibinfo {volume} {99}},\ \bibinfo {pages}
  {164} (\bibinfo {year} {2008})}\BibitemShut {NoStop}%
\bibitem [{\citenamefont {Park}\ \emph {et~al.}(2018)\citenamefont {Park},
  \citenamefont {Jo}, \citenamefont {Lee},\ and\ \citenamefont
  {Kim}}]{park2018generalized}%
  \BibitemOpen
  \bibfield  {author} {\bibinfo {author} {\bibfnamefont {H.~J.}\ \bibnamefont
  {Park}}, \bibinfo {author} {\bibfnamefont {W.~S.}\ \bibnamefont {Jo}},
  \bibinfo {author} {\bibfnamefont {S.~H.}\ \bibnamefont {Lee}},\ and\ \bibinfo
  {author} {\bibfnamefont {B.~J.}\ \bibnamefont {Kim}},\ }\bibfield  {title}
  {\bibinfo {title} {Generalized gravity model for human migration},\
  }\href@noop {} {\bibfield  {journal} {\bibinfo  {journal} {New Journal of
  Physics}\ }\textbf {\bibinfo {volume} {20}},\ \bibinfo {pages} {093018}
  (\bibinfo {year} {2018})}\BibitemShut {NoStop}%
\bibitem [{\citenamefont {Prieto~Curiel}\ \emph {et~al.}(2018)\citenamefont
  {Prieto~Curiel}, \citenamefont {Pappalardo}, \citenamefont {Gabrielli},\ and\
  \citenamefont {Bishop}}]{prieto2018gravity}%
  \BibitemOpen
  \bibfield  {author} {\bibinfo {author} {\bibfnamefont {R.}~\bibnamefont
  {Prieto~Curiel}}, \bibinfo {author} {\bibfnamefont {L.}~\bibnamefont
  {Pappalardo}}, \bibinfo {author} {\bibfnamefont {L.}~\bibnamefont
  {Gabrielli}},\ and\ \bibinfo {author} {\bibfnamefont {S.~R.}\ \bibnamefont
  {Bishop}},\ }\bibfield  {title} {\bibinfo {title} {Gravity and scaling laws
  of city to city migration},\ }\href@noop {} {\bibfield  {journal} {\bibinfo
  {journal} {PloS one}\ }\textbf {\bibinfo {volume} {13}},\ \bibinfo {pages}
  {e0199892} (\bibinfo {year} {2018})}\BibitemShut {NoStop}%
\bibitem [{\citenamefont {Erlander}\ and\ \citenamefont
  {Stewart}(1990)}]{erlander1990gravity}%
  \BibitemOpen
  \bibfield  {author} {\bibinfo {author} {\bibfnamefont {S.}~\bibnamefont
  {Erlander}}\ and\ \bibinfo {author} {\bibfnamefont {N.~F.}\ \bibnamefont
  {Stewart}},\ }\href@noop {} {\emph {\bibinfo {title} {The gravity model in
  transportation analysis: theory and extensions}}},\ Vol.~\bibinfo {volume}
  {3}\ (\bibinfo  {publisher} {Vsp},\ \bibinfo {year} {1990})\BibitemShut
  {NoStop}%
\bibitem [{\citenamefont {Jung}\ \emph {et~al.}(2008)\citenamefont {Jung},
  \citenamefont {Wang},\ and\ \citenamefont {Stanley}}]{jung2008gravity}%
  \BibitemOpen
  \bibfield  {author} {\bibinfo {author} {\bibfnamefont {W.-S.}\ \bibnamefont
  {Jung}}, \bibinfo {author} {\bibfnamefont {F.}~\bibnamefont {Wang}},\ and\
  \bibinfo {author} {\bibfnamefont {H.~E.}\ \bibnamefont {Stanley}},\
  }\bibfield  {title} {\bibinfo {title} {Gravity model in the korean highway},\
  }\href@noop {} {\bibfield  {journal} {\bibinfo  {journal} {Europhysics
  Letters}\ }\textbf {\bibinfo {volume} {81}},\ \bibinfo {pages} {48005}
  (\bibinfo {year} {2008})}\BibitemShut {NoStop}%
\bibitem [{\citenamefont {Wang}\ \emph {et~al.}(2019)\citenamefont {Wang},
  \citenamefont {Ma}, \citenamefont {Jiang}, \citenamefont {Yan},\ and\
  \citenamefont {Zhou}}]{wang2019gravity}%
  \BibitemOpen
  \bibfield  {author} {\bibinfo {author} {\bibfnamefont {L.}~\bibnamefont
  {Wang}}, \bibinfo {author} {\bibfnamefont {J.-C.}\ \bibnamefont {Ma}},
  \bibinfo {author} {\bibfnamefont {Z.-Q.}\ \bibnamefont {Jiang}}, \bibinfo
  {author} {\bibfnamefont {W.}~\bibnamefont {Yan}},\ and\ \bibinfo {author}
  {\bibfnamefont {W.-X.}\ \bibnamefont {Zhou}},\ }\bibfield  {title} {\bibinfo
  {title} {Gravity law in the chinese highway freight transportation
  networks},\ }\href@noop {} {\bibfield  {journal} {\bibinfo  {journal} {EPJ
  Data Science}\ }\textbf {\bibinfo {volume} {8}},\ \bibinfo {pages} {1}
  (\bibinfo {year} {2019})}\BibitemShut {NoStop}%
\bibitem [{\citenamefont {Li}\ \emph {et~al.}(2021)\citenamefont {Li},
  \citenamefont {Gao}, \citenamefont {Luo}, \citenamefont {Yao}, \citenamefont
  {Chen}, \citenamefont {Shang}, \citenamefont {Jiang},\ and\ \citenamefont
  {Stanley}}]{li2021gravity}%
  \BibitemOpen
  \bibfield  {author} {\bibinfo {author} {\bibfnamefont {R.}~\bibnamefont
  {Li}}, \bibinfo {author} {\bibfnamefont {S.}~\bibnamefont {Gao}}, \bibinfo
  {author} {\bibfnamefont {A.}~\bibnamefont {Luo}}, \bibinfo {author}
  {\bibfnamefont {Q.}~\bibnamefont {Yao}}, \bibinfo {author} {\bibfnamefont
  {B.}~\bibnamefont {Chen}}, \bibinfo {author} {\bibfnamefont {F.}~\bibnamefont
  {Shang}}, \bibinfo {author} {\bibfnamefont {R.}~\bibnamefont {Jiang}},\ and\
  \bibinfo {author} {\bibfnamefont {H.~E.}\ \bibnamefont {Stanley}},\
  }\bibfield  {title} {\bibinfo {title} {Gravity model in dockless bike-sharing
  systems within cities},\ }\href@noop {} {\bibfield  {journal} {\bibinfo
  {journal} {Physical Review E}\ }\textbf {\bibinfo {volume} {103}},\ \bibinfo
  {pages} {012312} (\bibinfo {year} {2021})}\BibitemShut {NoStop}%
\bibitem [{\citenamefont {Masucci}\ \emph {et~al.}(2013)\citenamefont
  {Masucci}, \citenamefont {Serras}, \citenamefont {Johansson},\ and\
  \citenamefont {Batty}}]{masucci2013gravity}%
  \BibitemOpen
  \bibfield  {author} {\bibinfo {author} {\bibfnamefont {A.~P.}\ \bibnamefont
  {Masucci}}, \bibinfo {author} {\bibfnamefont {J.}~\bibnamefont {Serras}},
  \bibinfo {author} {\bibfnamefont {A.}~\bibnamefont {Johansson}},\ and\
  \bibinfo {author} {\bibfnamefont {M.}~\bibnamefont {Batty}},\ }\bibfield
  {title} {\bibinfo {title} {Gravity versus radiation models: On the importance
  of scale and heterogeneity in commuting flows},\ }\href@noop {} {\bibfield
  {journal} {\bibinfo  {journal} {Physical Review E}\ }\textbf {\bibinfo
  {volume} {88}},\ \bibinfo {pages} {022812} (\bibinfo {year}
  {2013})}\BibitemShut {NoStop}%
\bibitem [{\citenamefont {Kwon}\ \emph {et~al.}(2023)\citenamefont {Kwon},
  \citenamefont {Hong}, \citenamefont {Jung},\ and\ \citenamefont
  {Jo}}]{kwon2023multiple}%
  \BibitemOpen
  \bibfield  {author} {\bibinfo {author} {\bibfnamefont {O.-H.}\ \bibnamefont
  {Kwon}}, \bibinfo {author} {\bibfnamefont {I.}~\bibnamefont {Hong}}, \bibinfo
  {author} {\bibfnamefont {W.-S.}\ \bibnamefont {Jung}},\ and\ \bibinfo
  {author} {\bibfnamefont {H.-H.}\ \bibnamefont {Jo}},\ }\bibfield  {title}
  {\bibinfo {title} {Multiple gravity laws for human mobility within cities},\
  }\href@noop {} {\bibfield  {journal} {\bibinfo  {journal} {EPJ Data Science}\
  }\textbf {\bibinfo {volume} {12}},\ \bibinfo {pages} {57} (\bibinfo {year}
  {2023})}\BibitemShut {NoStop}%
\bibitem [{\citenamefont {Tamura}\ \emph {et~al.}(2012)\citenamefont {Tamura},
  \citenamefont {Miura}, \citenamefont {Takayasu}, \citenamefont {Takayasu},
  \citenamefont {Kitajima},\ and\ \citenamefont {Goto}}]{tamura2012estimation}%
  \BibitemOpen
  \bibfield  {author} {\bibinfo {author} {\bibfnamefont {K.}~\bibnamefont
  {Tamura}}, \bibinfo {author} {\bibfnamefont {W.}~\bibnamefont {Miura}},
  \bibinfo {author} {\bibfnamefont {M.}~\bibnamefont {Takayasu}}, \bibinfo
  {author} {\bibfnamefont {H.}~\bibnamefont {Takayasu}}, \bibinfo {author}
  {\bibfnamefont {S.}~\bibnamefont {Kitajima}},\ and\ \bibinfo {author}
  {\bibfnamefont {H.}~\bibnamefont {Goto}},\ }\bibfield  {title} {\bibinfo
  {title} {Estimation of flux between interacting nodes on huge inter-firm
  networks},\ }in\ \href@noop {} {\emph {\bibinfo {booktitle} {International
  Journal of Modern Physics: Conference Series}}},\ Vol.~\bibinfo {volume}
  {16}\ (\bibinfo {organization} {World Scientific},\ \bibinfo {year} {2012})\
  pp.\ \bibinfo {pages} {93--104}\BibitemShut {NoStop}%
\bibitem [{\citenamefont {Wang}\ \emph {et~al.}(2021)\citenamefont {Wang},
  \citenamefont {Yan},\ and\ \citenamefont {Wu}}]{wang2021free}%
  \BibitemOpen
  \bibfield  {author} {\bibinfo {author} {\bibfnamefont {H.}~\bibnamefont
  {Wang}}, \bibinfo {author} {\bibfnamefont {X.-Y.}\ \bibnamefont {Yan}},\ and\
  \bibinfo {author} {\bibfnamefont {J.}~\bibnamefont {Wu}},\ }\bibfield
  {title} {\bibinfo {title} {Free utility model for explaining the social
  gravity law},\ }\href@noop {} {\bibfield  {journal} {\bibinfo  {journal}
  {Journal of Statistical Mechanics: Theory and Experiment}\ }\textbf {\bibinfo
  {volume} {2021}},\ \bibinfo {pages} {033418} (\bibinfo {year}
  {2021})}\BibitemShut {NoStop}%
\bibitem [{\citenamefont {de~Dios~Ort{\'u}zar}\ and\ \citenamefont
  {Willumsen}(2024)}]{de2024modelling}%
  \BibitemOpen
  \bibfield  {author} {\bibinfo {author} {\bibfnamefont {J.}~\bibnamefont
  {de~Dios~Ort{\'u}zar}}\ and\ \bibinfo {author} {\bibfnamefont {L.~G.}\
  \bibnamefont {Willumsen}},\ }\href@noop {} {\emph {\bibinfo {title}
  {Modelling transport}}}\ (\bibinfo  {publisher} {John wiley \& sons},\
  \bibinfo {year} {2024})\BibitemShut {NoStop}%
\bibitem [{\citenamefont {Tamura}\ \emph {et~al.}(2018)\citenamefont {Tamura},
  \citenamefont {Takayasu},\ and\ \citenamefont
  {Takayasu}}]{tamura2018diffusion}%
  \BibitemOpen
  \bibfield  {author} {\bibinfo {author} {\bibfnamefont {K.}~\bibnamefont
  {Tamura}}, \bibinfo {author} {\bibfnamefont {H.}~\bibnamefont {Takayasu}},\
  and\ \bibinfo {author} {\bibfnamefont {M.}~\bibnamefont {Takayasu}},\
  }\bibfield  {title} {\bibinfo {title} {Diffusion-localization transition
  caused by nonlinear transport on complex networks},\ }\href@noop {}
  {\bibfield  {journal} {\bibinfo  {journal} {Scientific reports}\ }\textbf
  {\bibinfo {volume} {8}},\ \bibinfo {pages} {5517} (\bibinfo {year}
  {2018})}\BibitemShut {NoStop}%
\bibitem [{\citenamefont {Koike}\ \emph {et~al.}(2022)\citenamefont {Koike},
  \citenamefont {Takayasu},\ and\ \citenamefont
  {Takayasu}}]{koike2022diffusion}%
  \BibitemOpen
  \bibfield  {author} {\bibinfo {author} {\bibfnamefont {H.}~\bibnamefont
  {Koike}}, \bibinfo {author} {\bibfnamefont {H.}~\bibnamefont {Takayasu}},\
  and\ \bibinfo {author} {\bibfnamefont {M.}~\bibnamefont {Takayasu}},\
  }\bibfield  {title} {\bibinfo {title} {Diffusion-localization transition
  point of gravity type transport model on regular ring lattices and bethe
  lattices},\ }\href@noop {} {\bibfield  {journal} {\bibinfo  {journal}
  {Journal of Statistical Physics}\ }\textbf {\bibinfo {volume} {186}},\
  \bibinfo {pages} {44} (\bibinfo {year} {2022})}\BibitemShut {NoStop}%
\bibitem [{\citenamefont {Koike}\ \emph {et~al.}(2024)\citenamefont {Koike},
  \citenamefont {Takayasu},\ and\ \citenamefont
  {Takayasu}}]{koike2024bifurcation}%
  \BibitemOpen
  \bibfield  {author} {\bibinfo {author} {\bibfnamefont {H.}~\bibnamefont
  {Koike}}, \bibinfo {author} {\bibfnamefont {H.}~\bibnamefont {Takayasu}},\
  and\ \bibinfo {author} {\bibfnamefont {M.}~\bibnamefont {Takayasu}},\
  }\bibfield  {title} {\bibinfo {title} {Bifurcation and hysteresis in a
  nonlinear transport model on network motifs},\ }\href@noop {} {\bibfield
  {journal} {\bibinfo  {journal} {Physical Review Research}\ }\textbf {\bibinfo
  {volume} {6}},\ \bibinfo {pages} {013059} (\bibinfo {year}
  {2024})}\BibitemShut {NoStop}%
\bibitem [{\citenamefont {Tamura}\ \emph {et~al.}(2015)\citenamefont {Tamura},
  \citenamefont {Takayasu},\ and\ \citenamefont
  {Takayasu}}]{tamura2015extraction}%
  \BibitemOpen
  \bibfield  {author} {\bibinfo {author} {\bibfnamefont {K.}~\bibnamefont
  {Tamura}}, \bibinfo {author} {\bibfnamefont {H.}~\bibnamefont {Takayasu}},\
  and\ \bibinfo {author} {\bibfnamefont {M.}~\bibnamefont {Takayasu}},\
  }\bibfield  {title} {\bibinfo {title} {Extraction of conjugate main-stream
  structures from a complex network flow},\ }\href@noop {} {\bibfield
  {journal} {\bibinfo  {journal} {Physical Review E}\ }\textbf {\bibinfo
  {volume} {91}},\ \bibinfo {pages} {042815} (\bibinfo {year}
  {2015})}\BibitemShut {NoStop}%
\bibitem [{Note1()}]{Note1}%
  \BibitemOpen
  \bibinfo {note} {The eigenvalues of the Jacobian of the model \protect \eqref
  {Eq:1} at uniform state are exactly calculated in Ref. \cite
  {koike2022diffusion}, and by taking continuum limit $N \rightarrow \infty $
  allows us to discuss the dispersion relation. See Section 1 of the
  Supplementary Material for details.}\BibitemShut {Stop}%
\bibitem [{\citenamefont {Shimada}\ and\ \citenamefont
  {Nagashima}(1979)}]{shimada1979numerical}%
  \BibitemOpen
  \bibfield  {author} {\bibinfo {author} {\bibfnamefont {I.}~\bibnamefont
  {Shimada}}\ and\ \bibinfo {author} {\bibfnamefont {T.}~\bibnamefont
  {Nagashima}},\ }\bibfield  {title} {\bibinfo {title} {A numerical approach to
  ergodic problem of dissipative dynamical systems},\ }\href@noop {} {\bibfield
   {journal} {\bibinfo  {journal} {Progress of theoretical physics}\ }\textbf
  {\bibinfo {volume} {61}},\ \bibinfo {pages} {1605} (\bibinfo {year}
  {1979})}\BibitemShut {NoStop}%
\bibitem [{Note2()}]{Note2}%
  \BibitemOpen
  \bibinfo {note} {Only one fixed point is shown in Fig. \ref {Fig:2}, but
  there are other fixed points owing to symmetry.}\BibitemShut {Stop}%
\bibitem [{\citenamefont {Hofbauer}\ and\ \citenamefont
  {Sigmund}(1998)}]{hofbauer1998evolutionary}%
  \BibitemOpen
  \bibfield  {author} {\bibinfo {author} {\bibfnamefont {J.}~\bibnamefont
  {Hofbauer}}\ and\ \bibinfo {author} {\bibfnamefont {K.}~\bibnamefont
  {Sigmund}},\ }\href@noop {} {\emph {\bibinfo {title} {Evolutionary games and
  population dynamics}}}\ (\bibinfo  {publisher} {Cambridge university press},\
  \bibinfo {year} {1998})\BibitemShut {NoStop}%
\bibitem [{\citenamefont {Chawanya}(1995)}]{chawanya1995new}%
  \BibitemOpen
  \bibfield  {author} {\bibinfo {author} {\bibfnamefont {T.}~\bibnamefont
  {Chawanya}},\ }\bibfield  {title} {\bibinfo {title} {A new type of irregular
  motion in a class of game dynamics systems},\ }\href@noop {} {\bibfield
  {journal} {\bibinfo  {journal} {Progress of Theoretical Physics}\ }\textbf
  {\bibinfo {volume} {94}},\ \bibinfo {pages} {163} (\bibinfo {year}
  {1995})}\BibitemShut {NoStop}%
\bibitem [{\citenamefont {May}\ and\ \citenamefont
  {Leonard}(1975)}]{may1975nonlinear}%
  \BibitemOpen
  \bibfield  {author} {\bibinfo {author} {\bibfnamefont {R.~M.}\ \bibnamefont
  {May}}\ and\ \bibinfo {author} {\bibfnamefont {W.~J.}\ \bibnamefont
  {Leonard}},\ }\bibfield  {title} {\bibinfo {title} {Nonlinear aspects of
  competition between three species},\ }\href@noop {} {\bibfield  {journal}
  {\bibinfo  {journal} {SIAM journal on applied mathematics}\ }\textbf
  {\bibinfo {volume} {29}},\ \bibinfo {pages} {243} (\bibinfo {year}
  {1975})}\BibitemShut {NoStop}%
\bibitem [{\citenamefont {Pomeau}\ and\ \citenamefont
  {Manneville}(1980)}]{pomeau1980intermittent}%
  \BibitemOpen
  \bibfield  {author} {\bibinfo {author} {\bibfnamefont {Y.}~\bibnamefont
  {Pomeau}}\ and\ \bibinfo {author} {\bibfnamefont {P.}~\bibnamefont
  {Manneville}},\ }\bibfield  {title} {\bibinfo {title} {Intermittent
  transition to turbulence in dissipative dynamical systems},\ }\href@noop {}
  {\bibfield  {journal} {\bibinfo  {journal} {Communications in Mathematical
  Physics}\ }\textbf {\bibinfo {volume} {74}},\ \bibinfo {pages} {189}
  (\bibinfo {year} {1980})}\BibitemShut {NoStop}%
\bibitem [{\citenamefont {Elaskar}\ and\ \citenamefont {del
  R{\'\i}o}(2023)}]{elaskar2023review}%
  \BibitemOpen
  \bibfield  {author} {\bibinfo {author} {\bibfnamefont {S.}~\bibnamefont
  {Elaskar}}\ and\ \bibinfo {author} {\bibfnamefont {E.}~\bibnamefont {del
  R{\'\i}o}},\ }\bibfield  {title} {\bibinfo {title} {Review of chaotic
  intermittency},\ }\href@noop {} {\bibfield  {journal} {\bibinfo  {journal}
  {Symmetry}\ }\textbf {\bibinfo {volume} {15}},\ \bibinfo {pages} {1195}
  (\bibinfo {year} {2023})}\BibitemShut {NoStop}%
\bibitem [{\citenamefont {Puu}\ and\ \citenamefont
  {Puu}(1991)}]{puu1991nonlinear}%
  \BibitemOpen
  \bibfield  {author} {\bibinfo {author} {\bibfnamefont {T.}~\bibnamefont
  {Puu}}\ and\ \bibinfo {author} {\bibfnamefont {T.}~\bibnamefont {Puu}},\
  }\href@noop {} {\emph {\bibinfo {title} {Nonlinear economic dynamics}}}\
  (\bibinfo  {publisher} {Springer},\ \bibinfo {year} {1991})\BibitemShut
  {NoStop}%
\bibitem [{\citenamefont {Puu}(2013)}]{puu2013attractors}%
  \BibitemOpen
  \bibfield  {author} {\bibinfo {author} {\bibfnamefont {T.}~\bibnamefont
  {Puu}},\ }\href@noop {} {\emph {\bibinfo {title} {Attractors, bifurcations,
  \& chaos: Nonlinear phenomena in economics}}}\ (\bibinfo  {publisher}
  {Springer Science \& Business Media},\ \bibinfo {year} {2013})\BibitemShut
  {NoStop}%
\bibitem [{\citenamefont {Barnett}\ \emph {et~al.}(2015)\citenamefont
  {Barnett}, \citenamefont {Serletis},\ and\ \citenamefont
  {Serletis}}]{barnett2015nonlinear}%
  \BibitemOpen
  \bibfield  {author} {\bibinfo {author} {\bibfnamefont {W.~A.}\ \bibnamefont
  {Barnett}}, \bibinfo {author} {\bibfnamefont {A.}~\bibnamefont {Serletis}},\
  and\ \bibinfo {author} {\bibfnamefont {D.}~\bibnamefont {Serletis}},\
  }\bibfield  {title} {\bibinfo {title} {Nonlinear and complex dynamics in
  economics},\ }\href@noop {} {\bibfield  {journal} {\bibinfo  {journal}
  {Macroeconomic Dynamics}\ }\textbf {\bibinfo {volume} {19}},\ \bibinfo
  {pages} {1749} (\bibinfo {year} {2015})}\BibitemShut {NoStop}%
\bibitem [{\citenamefont {San~Miguel}\ and\ \citenamefont
  {Toral}(2020)}]{san2020introduction}%
  \BibitemOpen
  \bibfield  {author} {\bibinfo {author} {\bibfnamefont {M.}~\bibnamefont
  {San~Miguel}}\ and\ \bibinfo {author} {\bibfnamefont {R.}~\bibnamefont
  {Toral}},\ }\bibfield  {title} {\bibinfo {title} {Introduction to the chaos
  focus issue on the dynamics of social systems},\ }\href@noop {} {\bibfield
  {journal} {\bibinfo  {journal} {Chaos: An Interdisciplinary Journal of
  Nonlinear Science}\ }\textbf {\bibinfo {volume} {30}} (\bibinfo {year}
  {2020})}\BibitemShut {NoStop}%
\bibitem [{\citenamefont {Gordon}(1992)}]{gordon1992chaos}%
  \BibitemOpen
  \bibfield  {author} {\bibinfo {author} {\bibfnamefont {T.~J.}\ \bibnamefont
  {Gordon}},\ }\bibfield  {title} {\bibinfo {title} {Chaos in social systems},\
  }\href@noop {} {\bibfield  {journal} {\bibinfo  {journal} {Technological
  Forecasting and Social Change}\ }\textbf {\bibinfo {volume} {42}},\ \bibinfo
  {pages} {1} (\bibinfo {year} {1992})}\BibitemShut {NoStop}%
\end{thebibliography}%

\end{document}